\documentclass[epsf,preprint,aps,12pt,showpacs,nofootinbib,tightenlines]{revtex4}

\usepackage[section]{placeins}
\usepackage{graphicx}% Include figure files
\usepackage{dcolumn}% Align table columns on decimal point
\usepackage{bm}% bold math

\usepackage{epsfig}
\usepackage{amssymb}
\usepackage{amsmath}
\usepackage{flafter}
\usepackage{array}

\usepackage[dvips,usenames]{color}

\newlength{\dinwidth}
\newlength{\dinmargin}
\begin{document}
\title{The Lee-Wick Fields out of Gravity}
\author{Feng Wu$^{1}$}
\email{fengwu@itp.ac.cn}
\author{Ming Zhong$^{2}$}
\email{zhongming@mail.tsinghua.edu.cn}

\affiliation{${}^{1}$ Kavli Institute for Theoretical Physics China (KITPC), Institute of
Theoretical Physics, Chinese Academy of Sciences, P.O. Box 2735, Beijing
100080, China}
\affiliation{${}^{2}$ Center for High Energy Physics, Department of Engineering Physics, Tsinghua University, Beijing 100084, China}
\begin{abstract}
We study the Maxwell-Einstein theory in the framework of effective field theories. We show that the modified one-loop renormalizable Lagrangian due to quantum gravitational effects contains a Lee-Wick vector field as an extra degree of freedom in the theory. Thus gravity provides a natural mechanism for the emergence of this exotic particle.  
\end{abstract}

%\pacs{  }%
\maketitle
The success of Quantum Electrodynamics (QED), a renormalizable $U(1)$ gauge field theory which respects Poincar$\acute{e}$ invariance, to account for almost all phenomena with precision down to subatomic scales has long made it the paradigm theory in particle physics. The idea that a sensible quantum field theory should be renormalizable  in order to make predictions also grew out of it. Renormalizability was considered to be an axiom when constructing the Standard Model in particle physics. Quantum corrections to a renormalizable theory will generate UV divergences only to the operators whose mass dimensions are less than five. This assures the predictiveness of the theory. General relativity, the theory of the gravitational force, on the other hand, is not renormalizable after quantization \cite{DeWitt}\cite{Veltman}. This is one of the reasons that general relativity was considered to be incompatible with quantum mechanics. In the quantized version of general relativity, one would not be able to reabsorb all the UV divergences into the coupling constants in the original Lagrangian. That is, new counter terms are needed at each order of perturbative calculations when trying to renormalize the theory. 

A modern point of view is that a non-renormalizable theory might be sensible and reliable predictions could still be made from it within the framework of effective field theories \cite{Polchinski}\cite{Kaplan}. From the value of the only dimensionful coupling constant $G$, Newton's constant, in the Hilbert-Einstein Lagrangian, one can see that gravitational effects are tiny at energies $E \ll M_{P} \sim 10^{19} GeV/c^2$. It makes sense to treat general relativity as a low energy effective field theory of some unknown fundamental theory and consider its quantum effects \cite{Donoghue}. The effects due to non-renormalizable terms are suppressed by inverse powers of $M_P$, the mass scale of new physics. 

In \cite{Robinson}, S. Robinson and F. Wilczek showed that gravitational corrections cause gauge couplings to run. More interestingly, the negative sign of the $\beta$ function due to gravitons in their result leads to the $U(1)$ gauge coupling constant running in the direction of asymptotic freedom at high energies. In contrast to pure Yang-Mills theories, quantum corrections do not introduce a mass gap to the pure Maxwell theory in a background flat spacetime and the $\beta$ function remains zero. This is because photons do not carry charge. The result of \cite{Robinson} shows that the one-loop quantum gravitational corrections will induce an anti-screening effect. Although the effect is tiny in the regions where perturbative calculations are reliable, it is of theoretical interest on its own. Later, A. Pietrykowski \cite{Pietrykowski} redid the calculations and found that in contrast to the conclusion in \cite{Robinson}, up to one-loop order the gravitational contributions to the $\beta$ function are indeed zero. The first relevant papers calculating the one-loop quantum gravitational effects on the Maxwell system are \cite{Deser}\cite{Deser1}.

In this work we will investigate the matter-free Maxwell-Einstein system. We focus our discussions on the gauge sector. The paper is organized into two parts. In the first part, we show by explicit calculations that up to one-loop order, quantum corrections to the gauge sector will induce a new type of divergences. Therefore, one needs to modify the Maxwell-Einstein Lagrangian and include one non-renormalizable term as an input for absorbing the divergent quantum correction. This modification is natural and consistent with the framework of effective field theories. The calculations provide a recheck of the results in \cite{Robinson} and \cite{Pietrykowski}. In the second part, we show that the modified one-loop renormalizable Maxwell-Einstein system contains a new degree of freedom in the theory, which is identified as the physical massive Lee-Wick photon \cite{Lee}\cite{Lee1}. Thus, the quantum gravitational effect suggests a mechanism to reveal this degree of freedom. 

We start with the Maxwell-Einstein theory. The action of the Maxwell-Einstein theory has the form:
\begin{equation}
S= - \int d^4x \sqrt{ - g} ({2\over \kappa^2} R + {1\over4} g^{\mu \lambda} g^{\nu \rho} F_{\mu \nu} F_{\lambda \rho} )
\label{action}
\end{equation}
where $\kappa^2 \equiv  32 \pi G$, $R$ is the Ricci scalar and $F_{\mu\nu}$ is the field strength tensor. We ignore the cosmological constant term in our discussions. The action (1) is invariant under general coordinate transformations and the $U(1)$ gauge transformation. Although photon-photon scattering cannot happen in the pure Maxwell theory and can only happen at the quantum level with the inclusion of matter fields, the second term in (1) shows that with gravitons as mediators this process can happen at the tree level. However, the effect is very weak since the amplitude is proportional to $\kappa^2 \sim {1\over M_{P}^2}$.  

In the following we use the background field method \cite{Abbott} and choose the background spacetime to be flat. Since the theory respects Poincar$\acute{e}$ invariance in flat spacetime, the gauge coupling constant is universal everywhere and the following discussion is without ambiguity. We assume the Maxwell equations are satisfied in the background spacetime. As is well known, using the background field method one is able to quantize a gauge field theory without losing explicit gauge invariance. In this framework, the $\beta$ function of the gauge coupling constant can be calculated from the renormalization constant of the background gauge field. This greatly simplifies the calculations, especially in the non-abelian and higher loop cases. Since gravitons do not carry gauge charge, this property still holds in the Maxwell-Einstein system. 

Fields $g_{\mu\nu}(x)$ and $ A_{\mu}(x)$ can be written as sums of background fields $(\eta_{\mu\nu}, \bar{A}_{\mu} )$ and quantum fluctuations $( h_{\mu\nu} , a_{\mu} )$:
\begin{equation}
g_{\mu\nu}(x)=\eta_{\mu\nu} + \kappa h_{\mu\nu}(x) , \;\;\;  A_{\mu}(x) = {1 \over \kappa} \bar{A}_{\mu}(x) + a_{\mu}(x).
\label{fields}
\end{equation}
Here $\eta_{\mu\nu}$ is the Minkowski metric. The convention for $\eta_{\mu\nu}$ in this paper has signature $(+,-,-,-)$. Rescaling the background gauge field in this way, one can easily show that in the action (1) the order $\kappa^{-2}$ terms are the classical Lagrangian and the order $\kappa^{-1}$ terms vanish after applying the equation of motion for the background field. In order to calculate the one-loop corrections, we need to know the order $\kappa^{0}$ terms. One can see from (2) and the action (1) that the $O(\kappa^0)$ terms are of the form $(hh)$, $(aa)$, $(ha\bar{A})$ and $(hh\bar{A}\bar{A})$. Expanding the action (1) in quantum fields, the $O(\kappa^{0})$ terms are:
\begin{align}
\int d^4 x [ & {1\over2}  \partial_{\nu} h_{\alpha\beta} P^{\alpha\beta\rho\sigma} \partial^{\nu} h _{\rho\sigma} - ( \partial^{\beta} h_{\beta \mu} - {1\over 2} \partial_{\mu} h )^2  - {1\over 2} ( \partial_{\mu} a_{\nu} )^2 + {1\over 2} (\partial_{\mu} a_{\nu} )( \partial^{\nu} a^{\mu} ) \nonumber \\
& -{1\over 2} h (\partial_{\mu} \bar{A}_{\nu} \partial^{\mu} a^{\nu} - \partial_{\nu} \bar{A}_{\mu} \partial^{\mu} a^{\nu} ) + h^{\beta \nu} (\partial_{\alpha} \bar{A}_{\nu} \partial^{\alpha} a_{\beta} +\partial_{\nu} \bar{A}_{\alpha} \partial_{\beta} a^{\alpha} - \partial_{\alpha} \bar{A}_{\nu} \partial_{\beta} a^{\alpha} - \partial_{\nu} \bar{A}_{\alpha} \partial^{\alpha} a_{\beta} ) ] \nonumber \\
& + (h h A A \;\;terms)
\label{a2}
\end{align}
where $P^{\alpha\beta\rho\sigma} \equiv {1\over 2} ( \eta^{\alpha\rho}\eta^{\beta\sigma} + \eta^{\alpha\sigma} \eta^{\beta\rho} - \eta^{\alpha\beta}\eta^{\rho\sigma} )$ and $h \equiv h_{\alpha}^{\alpha}$. 

To continue, one has to fix gauges. Adding the de Donder gauge fixing term ${1 \over \xi}( \partial^{\beta} h_{\beta \mu} - {1\over 2} \partial_{\mu} h )^2$ for the graviton field and the Lorentz gauge fixing term ${-1\over 2 \zeta} ( \partial^{\alpha} a_{\alpha} )^2$ for the photon field to the action, the propagators for the quantum graviton and the photon in momentum space are:
\begin{align}
& graviton: \;\;  {i \over p^2+ i \epsilon } [ P_{\alpha\beta\rho\sigma} - {(1-\xi) \over 2p^2 } ( p_{\alpha} p_{\rho} \eta_{\beta\sigma} + p_{\beta} p_{\rho} \eta_{\alpha\sigma} +p_{\alpha} p_{\sigma} \eta_{\beta\rho} + p_{\beta} p_{\sigma} \eta_{\alpha\rho})] \\
& photon: \;\; {-i \over q^2 + i \epsilon} [ \eta_{\alpha\beta} -(1- \zeta) { q_{\alpha} q_{\beta} \over q^2}].
\label{a2}
\end{align}

\begin{figure}[t]
\begin{center}
\includegraphics[width=10cm,clip=true,keepaspectratio=true]{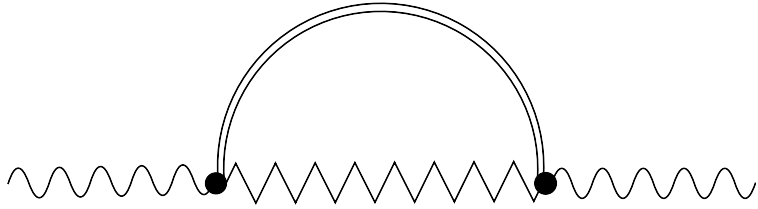}
\caption{\small The one-loop diagram generating the higher derivative term
. The internal wavy and double lines represent $a_{\mu}$ and $h_{\alpha\beta}$ fields respectively. The external lines are background photon fields.}
\end{center}\label{gr}
\end{figure}    

From the interaction terms on the second line of (3), it can be easily shown that the only non-trivial one-loop correction to the two point function of the photon field in the dimensional regularization scheme is the one shown in Fig. 1. The diagram is logarithmically divergent. All the tadpole diagrams vanish in this scheme. Note that we do not include ghost parts in (3) since they do not contribute to the one-loop order. We evaluate the integrals in $n=4-\epsilon$ dimensional spacetime to extract the singularities. The calculation is straightforward. One integral that is useful in evaluating this diagram is:
\begin{equation}
\int {d^n q \over (2\pi)^{n}} {q^{\mu} q^{\nu} \over (p+q)^2 q^2} = f(p^2, n) (-p^2 \eta^{\mu\nu} + n p^{\mu} p^{\nu})
\label{int}
\end{equation}
where $f(p^2,n) = { i \over (4 \pi)^{n/2}} {\Gamma({n\over 2}) \Gamma({n \over 2}-1) \over 2 \Gamma(n)} \Gamma(2-{n\over 2}) (-p^2)^{{n\over2}-2} $. Note that in $n$ dimensional spacetime, $P^{\alpha\beta\rho\sigma}$ in (4) takes the form ${1\over 2} ( \eta^{\alpha\rho}\eta^{\beta\sigma} + \eta^{\alpha\sigma} \eta^{\beta\rho} - {2\over n-2} \eta^{\alpha\beta}\eta^{\rho\sigma} )$. By naive power counting, the diagram is quadratically divergent. However, explicit calculations show that all quadratically divergent parts cancel each other and the result is logarithmically divergent. This is an example showing that for some Feynman integrals the superficial degree of divergence is lower than the one based on power counting. In QED, it is known that the one loop Fermion self-energy is logarithmically rather than linearly divergent because of the Lorentz invariance. Here the diagram is logarithmically rather than quadratically divergent because of the requirement of gauge invariance. By explicit calculation, it can be shown that the correction so obtained in the framework of background field method is gauge independent. The result is:
\begin{equation}
\kappa^2({n-5 \over 2}) f(p^2,n) p^2 (p^2 \eta^{\mu\nu} - p^{\mu} p^{\nu})={i \kappa^2 \over 192 \pi^2} {1 \over \epsilon} p^2 (p^2 \eta^{\mu\nu}- p^{\mu} p^{\nu}) + \; [finite\;\; part].
\label{result}
\end{equation}
This one-loop contribution to the photon-photon correlation function shows that in contrast to renormalizable theories, the correction due to gravitons generates a new type of divergence which cannot be absorbed in the Maxwell-Einstein action. In flat spacetime, the term $i(p_{\mu}p_{\nu} - p^2 \eta_{\mu\nu})$ in the truncated photon-photon correlation function corresponds to the operator $-{1\over 4} F_{\mu\nu} F^{\mu\nu}$. Similarly, one can show that the term $ip^2(p_{\mu}p_{\nu} - p^2 \eta_{\mu\nu})$ corresponds to the dimension six operator $-{1\over 2} \partial _{\mu}F^{\mu\nu} \partial^{\rho}F_{\rho\nu}$. Thus, the result in (7) shows that the new counter term needed is of the form $\partial_{\mu} F^{\mu\nu} \partial^{\rho} F_{\rho\nu}$. This is the leading higher derivative term allowed by symmetries. Without renormalizability as an axiom, there is no reason to include in the input Lagrangian only operators with mass dimension less than five. And we show here that if one unnaturally neglects this term in the action (1), the theory will lack its predictiveness at one-loop order. Since this is the only quantum correction to the gauge field, the dimension four operator $F^{\mu\nu} F_{\mu\nu}$ does not get any correction due to gravitons and the $\beta$ function remains zero.

Now let us consider the modified Maxwell-Einstein theory with the required higher derivative term. The  action has the form:
\begin{equation}
- \int d^4x \sqrt{ - g} ({1\over \kappa^2} R + {1\over4} g^{\mu \lambda} g^{\nu \rho} F_{\mu \nu} F_{\lambda \rho}-{a_{1} \over M_{P}^2} D_{\mu} F^{\mu\nu} D^{\rho} F_{\rho\nu} ) 
\label{modlag}
\end{equation}  
where $a_{1}$ is a dimensionless parameter and $D_{\mu}$ is the spacetime covariant derivative. Based on the previous discussion the origin of the last term is clear. The existence of gravity naturally provides a mechanism to generate this non-renormalizable term. With this term the gauge sector in the theory is one-loop renormalizable. In the limit $M_{P} \rightarrow \infty$, gravitons decouple from the theory and the action (8) goes back to the pure Maxwell theory in flat spacetime. 

Compare with the action (1), the modified Maxwell-Einstein theory has one additional free parameter, $a_{1}$. In the following, we focus our discussions on the case where $a_{1}$ is positive.

Let us consider again the background spacetime to be a Minkowski one and focus our discussion on the gauge sector. The part of the action we are interested in in order to find the propagator is quadratic in terms of the gauge fields:
\begin{equation}
- \int d^4x ({1\over4} g^{\mu \lambda} g^{\nu \rho} F_{\mu \nu} F_{\lambda \rho}-{a_{1} \over M_{P}^2} \partial_{\mu} F^{\mu\nu} \partial^{\rho} F_{\rho\nu} ) .
\label{propagator}
\end{equation}  
With the same Lorentz gauge fixing term as before, the propagator in the background spacetime is:
\begin{equation}
{-i \over q^2 - {a_{1} \over M_{P}^{2}}q^4+i \epsilon} [ \eta_{\mu\nu} -  {q_{\mu}q_{\nu} \over q^2}+ \zeta (1-a_{1} {q^2 \over M_{P}^2} ){q_{\mu} q_{\nu} \over q^2} ].
\label{propagator1}
\end{equation}  
For positive $a_1$, the propagator contains two poles. One corresponds to the massless photon and the other one at $q^2={M_{P}^2 \over a_{1}}$ corresponds to the Lee-Wick field, a massive spin one vector field \cite{Lee}\cite{Lee1}. Thus, the gravity reveals one extra degree of freedom. 

At low energies, the effects from the massless photon fields dominate over the Lee-Wick field contributions. However, making the replacement
\begin{equation}
{1\over q^2 - {a_{1} \over M_{P}^{2}}q^4} =  {1\over q^2} - {1\over q^2- {M_{P}^2\over a_{1}}},  
\label{replace}
\end{equation}  
suggests that the massive Lee-Wick field can play the role of the regulator in Pauli-Villars regularization. It was first studied by Lee and Wick in attempts to construct another version of QED with finite mass, charge and wave function renormalizations. This new degree of freedom was argued to be a physical resonance and unitarity is preserved despite the wrong sign in (11). As a result, one is left with a theory containing both massless and massive spin one fields. Similar to photons, the massive Lee-Wick vector fields also serve as mediators of the electromagnetic interactions. At low energies, we will not see a physical Lee-Wick particle, and the structure of the gauge field propagator indicates that the electromagnetic potential deviates from the Coulomb potential. We will not attempt to study its scaling behavior here.  

Recently, Grinstein, O'Connell and Wise \cite{Grinstein} have used this idea and shown that with the inclusion of higher derivative terms in each sector of the Standard Model, the extended model is free of quadratic divergences. Thus, including higher derivative terms provides a new way to stabilize the mass of the Higgs particle. Here we consider the simplified case with only $U(1)$ gauge bosons and gravitons to show that the higher derivative term is inevitable with its origin coming from gravitational corrections. In the case of scalar fields $\phi$ interacting with gravitons, a dimension six operator of the form $ \sim (D^2 \phi)^{\dagger}(D^2\phi)$ is required as an input in the original Lagrangian in order to renormalize the theory to one-loop order \cite{Veltman}. The modified version of the theory with this dimension six operator contains the Lee-Wick scalars as regulators and is free of quadratic divergences. 

From the point of view of effective field theories, non-renormalizable terms are inversely proportional to powers of the energy scale of new physics. No matter how weak the gravitational effect is, as long as it exists, it naturally provides an origin of the required higher derivative terms and this in turn reveals the Lee-Wick fields as extra degrees of freedom. 

We would like to emphasize that the same mechanism happens in the case of non-abelian gauge bosons and matter fields. Since gravitons do not carry gauge charges, the same results can be applied directly to the non-abelian case.

\date{\today}

\section*{Acknowledgements}

F.W. is supported in part by the Project of Knowledge Innovation Program (PKIP) of Chinese Academy of Sciences (CAS).

\end{document}